\documentclass{emulateapj}
\shorttitle{Turbulence and conduction in clusters}
\shortauthors{Ruszkowski \& Oh}

\newcommand{\brunt}{Brunt-V\"ais\"al\"a \,}
\def\gsim{\;\rlap{\lower 2.5pt
 \hbox{$\sim$}}\raise 1.5pt\hbox{$>$}\;}
\def\lsim{\;\rlap{\lower 2.5pt
   \hbox{$\sim$}}\raise 1.5pt\hbox{$<$}\;}

\begin{document}

\title{Shaken and stirred: conduction and turbulence in clusters of galaxies} 

\author{M. Ruszkowski\altaffilmark{1}}
\affil{Department of Astronomy, University of Michigan, 500 Church Street, Ann Arbor, MI 48109, USA; \\
e-mail: mateuszr@umich.edu (MR)}
\affil{The Michigan Center for Theoretical Physics, 3444 Randall Lab, 450 Church St, Ann Arbor, MI 48109, USA}
\author{S. Peng Oh \altaffilmark{1}}
\affil{Department of Physics, University of California, Santa Barbara, CA 93106, USA; \\
e-mail: peng@physics.ucsb.edu (SPO)}
\altaffiltext{1}{work performed in part while on leave at Max Planck Institute for Astrophysics, 
Karl-Schwarzschild-Str. 1, 85741 Garching, Germany and at Institute of Astronomy, Madingley Road, Cambridge CB3 0HA, United Kingdom}

\begin{abstract}
Uninhibited radiative cooling in clusters of galaxies would lead to excessive mass accretion rates contrary to observations.
One of the key proposals to offset radiative energy losses is thermal conduction from outer, hotter layers of 
cool core clusters to their centers. However, thermal conduction is sensitive to magnetic field topology. 
In cool-core clusters where temperature decreases inwards, the heat buoyancy instability (HBI) leads to magnetic fields ordered preferentially 
in the direction perpendicular to that of gravity, which significantly reduces the level of conduction below the classical 
Spitzer-Braginskii 
value. However, the cluster cool cores are rarely in perfect hydrostatic equilibrium. Sloshing motions due to minor mergers and
stirring motions induced by cluster galaxies or active galactic nuclei
(AGN) can significantly perturb the gas. The turbulent cascade can then affect the topology of the magnetic field and the effective 
level of thermal conduction.
We perform three-dimensional adaptive mesh refinement magnetohydrodynamical (MHD) simulations of the effect of turbulence on the properties of 
the anisotropic thermal conduction in cool core clusters.
We show that very weak subsonic motions,
well within observational constraints, 
can randomize the magnetic field 
and significantly boost effective thermal conduction beyond the saturated values expected in the pure unperturbed HBI case. 
We 
find
that the turbulent motions can essentially restore the conductive heat flow to the cool core to level comparable to the 
theoretical maximum of $\sim 1/3$ Spitzer for a highly tangled field. 
Runs with radiative cooling show that the cooling catastrophe can be averted and the cluster core stabilized; 
however, this conclusion may depend on the central gas density.
Above a critical Froude number, these same turbulent motions also eliminate the tangential bias in the velocity and magnetic field 
that is otherwise induced by the trapped $g$-modes. 
Our results can be tested with future radio polarization measurements, and have implications for efficient metal dispersal in clusters. \\
\end{abstract}

\keywords{conduction -- cooling flows -- galaxies: clusters: general -- galaxies: active -- instabilities -- X-rays: galaxies: clusters}

\section{Introduction}
\label{section:intro}

X-ray clusters show a strong central surface brightness peak, and many of them have short central cooling times. However, {\it Chandra} and {\it XMM-Newton} observations have shown that actual gas cooling rates fall significantly below classical values from a standard cooling flow model; emission lines such as Fe XVII expected from low temperature gas are not observed \citep{peterson06}. This suggests that besides quasi-hydrostatic equilibrium, clusters might also be in quasi-thermal equilibrium, with some form of heating balancing cooling. Many forms of heating have been postulated.
Hybrid models can account for the observations; leading contenders generally involve some combination of AGN heating and thermal conduction. The remarkable discovery by {\it Chandra} of AGN blown bubbles in clusters has triggered a widespread renaissance in the study of AGN feedback (see \citet{mcnamara07} for a recent review). It has been suggested that the AGN feedback {\it alone}, 
generally fails to distribute heat in the  spatially distributed fashion required to offset cooling (e.g., Vernaleo \& Reynolds 2006), both in its radial dependence and solid angle coverage. However, recent simulations by Br{\"u}ggen \& Scannapieco (2009) involving 
subgrid model for Rayleigh-Taylor-driven turbulence show that AGN heating can self-regulate cool cores.
In these non-MHD simulations the subgrid turbulence model is crucial to simulate the evolution of the cool core.
In a realistic situation, the ability to self-regulate  
undoubtely depends on a range of the mechanisms used to transmit the energy from the AGN blown bubbles to the ICM (e.g., dissipative sound waves, $pdV$ work, cosmic rays, suppression of both the Rayleigh-Taylor instability and the mixing of the bubble material by (even weak) magnetic fields). These issues are a matter of intense debate. At the same time, it is a remarkable fact that if one simply uses observed temperature profiles in clusters to construct the Spitzer conductive flux from the cluster outskirts, it is very nearly equal to that required to balance the radiative cooling rate as indicated by the observed X-ray surface brightness profile, for some reasonable fraction $f\sim 0.3$ of the Spitzer value (e.g., Fig. 17 of Peterson \& Fabian 2006). There is no reason in principle why such close agreement should exist, and it is a tantalizing hint that nature somehow ``knows'' about Spitzer conductivity. Models which invoke both mechanisms are able to reproduce observed temperature and density profiles without any fine-tuning of feedback or conduction parameters \citep{ruszkowski02,guo08a}, unlike conduction-only models \citep{bertschinger86,zakamska03}.

Nonetheless, even if no fine-tuning of conductivity is required, it is unsatisfying to invoke an ad-hoc suppression factor $f$; the anisotropic transport of heat along field lines can be calculated from first principles. If the field lines are tangled and chaotic, then $f\sim 0.2$ might be appropriate \citep{narayan01}. However, recent work has uncovered instabilities arising in stratified plasmas, which rearrange the field lines on large scales, with strong implications for heat transport. Unlike convective instability, which only arises if entropy declines with radius, these instabilities depend on temperature gradients, and arise irrespective of the sign of the gradient. If $dT/dr < 0$, as in the outer regions of clusters, the resulting magnetothermal instability (MTI; \citet{balbus00,parrish05,parrish07,parrish08}) drives field lines to become preferentially radial, substantially boosting conductivity to close to the Spitzer value, $f \sim 1$. On the other hand, if $dT/dr > 0$, as in the cores of cool-core clusters, then the resulting heat buoyancy instability (HBI; \citet{quataert08, parrish09}, Bogdanovi{\'c} et al. 2009) will drive field lines to become preferentially tangential, essentially shutting off radial conduction ($ f \ll 1$), and leading to catastrophic cooling in the cluster core. Thus, it would seem that a more careful calculation vitiates the ad-hoc assumptions of models which invoke conduction. 

However, this presumes that there are no competing mechanisms which also affect field line topology. One possibility which has been speculated upon in the literature \citep{ruszkowski07,guo08b,bogdanovic09} is AGN bubbles, which upon rising would leave radially orientated field lines in their wake. \citet{guo09} showed that time-variable conduction, perhaps triggered by AGN outbursts, could allow clusters to cycle between the cool core and non-cool core states. Another, perhaps more general mechanism is simply turbulent gas motions, which can be induced by mergers \citep{ascasibar06}, the motion of galaxies \citep{kim07,conroy08}, AGN outbursts \citep{mcnamara07}, or cosmic-ray driven convection \citep{chandran07,sharma09}\footnote{Note, however, that since the HBI is a buoyancy instability rather than a convective instability, turbulence cannot be self-consistently generated by the HBI itself, since the induced velocities are small \citep{bogdanovic09,parrish09}.}. Indeed, \citet{sharma09a} suggested that the amount of turbulence needed to avoid the stable end-state induced by the HBI could be estimated from the Richardson number \citep{turner73}, which is simply the ratio of the restoring buoyant force to the inertial $(\rho {\bf u}\cdot \nabla {\bf u})$ term. If one defines $Ri \equiv g r ( d \, {\rm ln}T/d \, {\rm ln} r)/\sigma^{2}$ for typical cluster conditions, the critical turbulent velocity is \citep{sharma09a}:

\begin{eqnarray}
\sigma \approx 135 \, {\rm km \, s^{-1}} {g_{-8}}^{1/2} r_{10}^{1/2}\left( \frac{d \, {\rm ln}T/d \, {\rm ln} r}{0.15} \right)^{1/2} \left( \frac{Ri_{c}}{0.25} \right)^{-1/2}\\ \nonumber
\end{eqnarray}

\noindent
where $g_{-8}$ is the gravitational acceleration in units of $10^{-8} \, {\rm cm^{2} s^{-1}}$, $r_{10}$ is a characteristic scale height in units of $10$ kpc, and $Ri_{c}$ is the critical Richardson number. $Ri_{c} \sim 1/4$ is typical for hydrodynamic flow; order unity differences might exist for the MHD case. Such levels of turbulence are easily seen in cosmological simulations, even in very relaxed clusters \citep{evrard90,norman99,nagai03,vazza09a, vazza09b}, 
presumably due to continuing gas accretion and the supersonic motions of galaxies through the ICM. Gas ``sloshing'' in clusters has been invoked to explain observed cold fronts \citep{ascasibar06}, and biases in X-ray mass profiles, particularly in the cluster core \citep{lau09}. In the future, direct measurement of turbulent velocities could best be performed by a high-spectral resolution imaging X-ray spectrometer \citep{rebusco08}, although early upper limits exist from {\it XMM-Newton} (\citet{sanders09}; for A1835, non-thermal broadening is less than $\sim 275 \, {\rm km \, s^{-1}}$). Besides the strong impact on conductive heat transport, the interaction of turbulence with conduction has testable implications for metal dispersal (which is strongly enhanced; \citep{sharma09}). With an instrument as as the Square Kilometer Array, the field topology itself can potentially be mapped out via the rotation measure of background radio sources.  

Our goal in this paper is to quantitatively examine for the first time if indeed turbulent motions can overwhelm the tangential field topology driven by the HBI (we defer the MTI dominated regime to future work), and allow thermal conduction at the level required to avert a cooling catastrophe. We begin with a MHD {\it FLASH} simulation of a cluster which in its unperturbed state suffers the HBI, consistent with previous work. 
We then simulate turbulence via a spectral forcing scheme that enables statistically stationary velocity fields (Eswaran \& Pope 1998). An important feature of our study is that even if the medium is stirred isotropically, the turbulent velocity field itself can be anisotropic, and interact in non-trivial ways with thermal conduction. This is because when the turbulent driving frequency falls below the \brunt frequency, {\it g}-modes can be excited; the trapped modes induce tangentially biased vortices, due to the restoring forces which act on vertical motion \citep{riley00}. Thus, even in the absence of the HBI, the excitation of $g$-modes can tangentially bias the magnetic field. In realistic clusters, $g$-modes also greatly enhance the efficacy of stirring, since trapped modes allow the turbulence to be volume-filling, although our present simulations already have volume-filling turbulence by design.  
The \brunt frequency, which determines where $g$-modes can be produced, depends both on the gravitational potential and the gas temperature and density profile. The latter can in turn be affected by heating and thermal conduction. Thus, the cluster potential, gas properties, and thermal conduction interact in often subtle ways. Overall, we find that turbulence can indeed restore conductive heat flow back to the values $f\sim 0.3$ expected for a highly tangled field, and required to stem a cooling flow. 

The organization of this paper is as follows. In \S\ref{section:turbulence}, we describe simple scaling relations to guide our intuition; in \S\ref{section:methods}, we describe our computational methods; in \S\ref{section:results}, our results; we then conclude in \S\ref{section:conclusions}. 

\section{Theoretical Expectations: Turbulence in a Stratified Medium}
\label{section:turbulence}

It is useful to begin by considering theoretical expectations for fluid motions in a stably stratified medium. We first review the case for a hydrodynamic fluid, which in clusters is stabilized by a strong entropy gradient. We then consider the MHD case where anisotropic conduction alters the picture. 

In the purely hydrodynamic case, the medium is convectively unstable if $d\, {\rm ln} S/d\, {\rm ln} r < 0$. On the other hand, if the medium is stably stratified with $d\, {\rm ln} S/d\, {\rm ln} r > 0$, a fluid element which is adiabatically displaced about its equilibrium position will experience a restoring force $f = \rho \ddot{r}= - \rho g (3/5) (d\, {\rm ln} S/d  r) \Delta r$, assuming it remains in pressure balance with its surroundings. This is the equation for simple harmonic motion, at frequency: 

\begin{eqnarray}
\omega_{\rm BV}^{2} = \frac{g}{r}\frac{3}{5} \frac{d\, {\rm ln} S}{d\, {\rm ln} r},\\ \nonumber
\end{eqnarray}

\noindent
the \brunt frequency. A formal WKB analysis reveals that the dispersion relation for such $g$-modes, where gravity is the restoring force, is: 

\begin{eqnarray}
\omega^{2} = \omega_{\rm BV}^{2} \frac{k_{\perp}^{2}}{k^{2}}\\ \nonumber
\end{eqnarray}

\noindent
where $k^{2}=k_{\perp}^{2} + k_{r}^{2}$. This immediately then implies that for oscillatory modes with real $k_{r}$, $\omega < \omega_{\rm BV}$; otherwise the modes have imaginary radial wavenumber and are evanescent. Physically, one can always achieve a low frequency response by making the mode progressively more tangential, but it is impossible to drive the system at frequencies higher than the maximum response frequency of $\omega_{\rm BV}$, corresponding to completely vertical oscillations. This thus implies that waves driven at frequencies $\omega < \omega_{\rm BV}$ can be resonantly excited, and will be trapped, reflected and focused inside the resonance radius where $\omega_{\rm BV}=\omega$. This has implications for how stirring can be amplified into volume-filling turbulence.  
                                                                                                                              
For our purposes, an even more interesting aspect of such $g$-modes is that they tend to be preferentially tangential, as has been widely confirmed in calculations of stellar pulsation \citep{cox80}, galaxy clusters \citep{lufkin95}, and the earth's atmosphere and oceans \citep{riley00}. Let us consider a simple order of magnitude argument for why this should be the case. For simplicity, consider an isothermal atmosphere. In steady state, and adopting the Boussinesq approximation (such that $\nabla \cdot {\bf v} =0$), the continuity equation reads ${\bf v} \cdot \nabla \rho =0$. Since the background density $\rho_{o}$ does not vary horizontally, the perturbed continuity equation is ${\bf v} \cdot \nabla \rho^{\prime} + v_{z} d\rho_{o}/dz = 0$. If $|{\bf v}| \sim U$, $|{ v_{z}}| \sim W$, and $L$ is a characteristic lengthscale, then:

\begin{eqnarray}
\rho^{\prime}  \sim \frac{W L}{U} \left| \frac{d \rho_{o}}{dz} \right| \\ \nonumber
\label{eq:OOM_continuity}
\end{eqnarray} 

\noindent
The buoyancy forces from stratification impose additional constraints. The vorticity evolution form of the momentum equation\footnote{Note that since $g$-modes generically generate vorticity, vorticity maps are a very useful marker of their presence.} for $g$-waves (i.e., assuming $\delta P/P \ll \delta \rho/\rho$) is \citep{lufkin95}:

\begin{eqnarray}
\frac{\partial {\bf (\delta \Omega)}}{\partial t} = i \frac{\rho^{\prime}}{\rho} ({\bf k} \times {\bf g}) \\ \nonumber
\end{eqnarray} 

\noindent
which gives to order of magnitude, 

\begin{eqnarray}
\rho^{\prime} \sim \rho_{o} \frac{U^{2}}{g L}.\\ \nonumber
\label{eq:OOM_vorticity}
\end{eqnarray}

\noindent
Combining equation (\ref{eq:OOM_continuity}) and (\ref{eq:OOM_vorticity}), we obtain:

\begin{eqnarray}
\frac{W}{U} \sim \frac{\rho_{o} U^{2}}{g L^{2} |d\rho/dz|} \sim \left(\frac{U}{\omega_{\rm BV} L} \right)^{2} \sim Fr^{2} \\ \nonumber
\label{eq:froude}
\end{eqnarray}

\noindent
where $Fr$ is the Froude number, which in this case compares the characteristic frequencies of inertial and gravitational (buoyancy) forces. $Ri\sim 1/Fr^{2}$ is often known as the Richardson number.  Equation (\ref{eq:froude}) reveals two important properties. When turbulence is absent or very weak, $Fr \ll 1$, motion is fundamentally horizontal, $W \ll U$: strong restoring buoyancy forces oppose motion in the vertical direction. However, stronger stirring, $Fr \gsim \mathcal{O}(1)$, can overwhelm such forces and re-establish isotropic turbulence. 

While all of the preceding discussion hold only for short wavelength modes in the WKB approximation, numerical simulations of non-linear evolution of global $g$-modes in galaxy clusters excited by orbiting galaxies show that they can still be very usefully interpreted in terms of the linear WKB analysis \citep{lufkin95}. 

How do these expectations change in the presence of anisotropic conduction? Here, behavior is governed by temperature rather than entropy gradients, and the atmosphere is buoyantly unstable regardless of the sign of $\nabla T$. For the HBI, the instability saturates once the magnetic field is perpendicular to $\nabla T$. Displaced fluid elements now experience a restoring force $f = \rho \ddot{r} = - \rho g (d\, {\rm ln} T/d  r) \Delta r$ \citep{sharma09a}, and hence oscillate about their equilibrium position at a frequency:

\begin{eqnarray}
(\omega_{\rm BV}^{\rm MHD})^{2} = \frac{g}{r}\frac{d\, {\rm ln} T}{d\, {\rm ln} r}, \\ \noindent
\end{eqnarray}

\noindent
which is typically $\sim$2 times smaller than $\omega_{\rm BV}$. Note that this assumes $t_{\rm cond} \sim L^{2}/\kappa \ll t_{\rm sc} \sim L/c_{s}$ (where $\kappa \sim v_{e} l_{e}$ is the conductivity coefficient) so that a displaced blob's temperature is determined by conductive rather than adiabatic cooling. This requires: 

\begin{eqnarray}
L < \sqrt{\frac{m_{p}}{m_{e}}} l_{e} \sim 200 \, {\rm kpc} \left( \frac{T}{4 \, {\rm keV}} \right)^{2} \left( \frac{n}{10^{-2} \, {\rm cm^{-3}}} \right)^{-1}\\ \nonumber
\end{eqnarray}

\noindent
which in any case is required by the WKB assumption, $L \ll r$. The dispersion relation is \citep{quataert08}:

\begin{eqnarray}
\omega^{2} = (\omega_{\rm BV}^{\rm MHD})^{2} \left[ (1-2 b_{z}^{2}) \frac{k_{\perp}^{2}}{k^{2}} + \frac{2 b_{x} b_{z} k_{x} k_{z}}{k^{2}} \right],\\ \noindent
\end{eqnarray}

\noindent
where $\hat{\bf z} = \hat{\bf r}$, $b_{x,z}=B_{x,z}/B$ and ${\bf B} = B_{x} \hat{\bf x} + B_{z} \hat{\bf z}$, without loss of generality. One can then solve the quadratic equation for $k_{z}$ to obtain the criterion for oscillatory solutions (real $k_{z}$), which yields the criterion $\omega \lsim (1-2 b_{z}^{2})^{1/2} \omega_{\rm BV}^{\rm MHD} \approx \omega_{\rm BV}^{\rm MHD}(1/\sqrt{3},1)$, corresponding to fully tangled (tangential) fields. Since temperature gradients are generally milder than entropy gradients, the criteria for $g$-modes is therefore marginally more stringent in the MHD case. The discussion of tangentially biased motion carries over with equal force to the MHD case, except for the change in the nature of the restoring force, such that:

\begin{eqnarray}
\frac{W}{U} \sim \left(\frac{U}{\omega_{\rm BV}^{\rm MHD} L} \right)^{2} \sim (Fr^{\rm MHD})^{2}\\ \nonumber
\label{eq:Fr_MHD} 
\end{eqnarray}

\noindent
so that if $Fr^{\rm MHD} \gsim \mathcal{O}(1)$, turbulent motions can isotropize fluid motions (and thus the magnetic field). 

We now proceed to explicitly test these notions in 3D, MHD simulations. It is important to note that $\omega_{\rm BV}^{\rm MHD}$ itself can be appreciably altered by gas motions. For instance, if thermal conduction is restored by stirring, heat transfer acts to reduce the temperature gradient and hence $\omega_{\rm BV}^{\rm MHD}$, further increasing $Fr^{\rm MHD}$. In the limit where efficient conduction causes the cluster to become isothermal, $\omega_{\rm BV}^{\rm MHD} \rightarrow 0$ and fluid elements are neutrally buoyant, with no restoring force upon displacement. Such a situation may indeed prevail in clusters which have fairly flat temperature profiles. In cool core clusters, the temperature profile likely depends on a complex interplay between turbulence, conduction, radiative cooling and AGN feedback. \\

\section{Methods}
\label{section:methods}

\subsection{Initial conditions for the gas}

For concreteness, we adopt simulation parameters appropriate for the well-observed $T_{\rm vir} \sim 5$ keV cool-core cluster A2199. Note that although we focus attention on the cluster core in the paper, we simulate the entire cluster in a large 1 Mpc box, to avoid sensitivity to assumed boundary conditions.\\
\indent
We model the dark matter distribution as a softened NFW profile of the form:
\begin{eqnarray}
\rho_{\rm DM}=\frac{M_{\rm o}/2\pi}{(r+r_{c})(r+r_{s})^{2}}, \\ \nonumber
\end{eqnarray}

\noindent
where $r_{c}=20$ kpc is the smoothing core radius and $r_{s}=390$ kpc is the usual NFW scale radius. The parameter $M_{o}=3.8\times 10^{14}M_{\odot}$
determines the cluster mass and is of the order of the cluster virial value. This density distribution corresponds to a 
gravitational acceleration:

\begin{eqnarray}
g & = & -\frac{GM_{\rm o}}{(r-r_{c})^{2}}\Bigl[-\frac{r_{s}(r_{s}-r_{c})}{r+r_{s}}+   \nonumber  \\ 
  &   & r_{s}(r_{s}-2r_{c})\ln\left(1+\frac{r}{r_{s}}\right)+r_{c}^{2}\ln\left(1+\frac{r}{r_{c}}\right)\Bigr]\\ \nonumber 
\end{eqnarray}

To model the gas, we adopted a power law entropy profile with a core, as seen in observations \citep{cavagnolo09}:

\begin{eqnarray}
K(r)=K_{0}+K_{1} \left( \frac{r}{r_{200}} \right)^{\alpha}, \\ \nonumber
\end{eqnarray}

\noindent
where entropy $K\equiv k_{B} T/n^{2/3}$. Here we adopt $\alpha=1.1$ and adjust $K_{0}$ and $K_{1}$ to match observed temperature and density profiles for A2199 ($K_{0}=k_{B}T_{0}/n_{0}^{2/3}$ where $T_{0},n_{0}$ are observed central temperature and density, and $K_{1}=K_{200}= k_{B}T_{200}/n_{200}^{2/3}$; we estimate from the virial theorem $k_{B} T_{200} \approx G M_{200} m_{p}/2 r_{200})$, and assume that the gas profile traces the dark matter profile in the outer regions. Given this entropy profile, we can solve for hydrostatic equilibrium: 

\begin{eqnarray}
\frac{d \, {\rm log}\rho}{d \, {\rm log} r} = - \frac{1}{\gamma} \left[ \frac{d \, {\rm log}S}{d \, {\rm log} r} + \frac{G M}{r k_{B} T} \right]\\ \nonumber
\end{eqnarray}

\noindent
where $k_{B}T=S n^{2/3}$, to obtain the temperature and density profiles. 
These initial conditions can be seen in Figure \ref{plot1}. 

\begin{figure}
\includegraphics[width=0.5\textwidth, height = 0.45\textwidth]{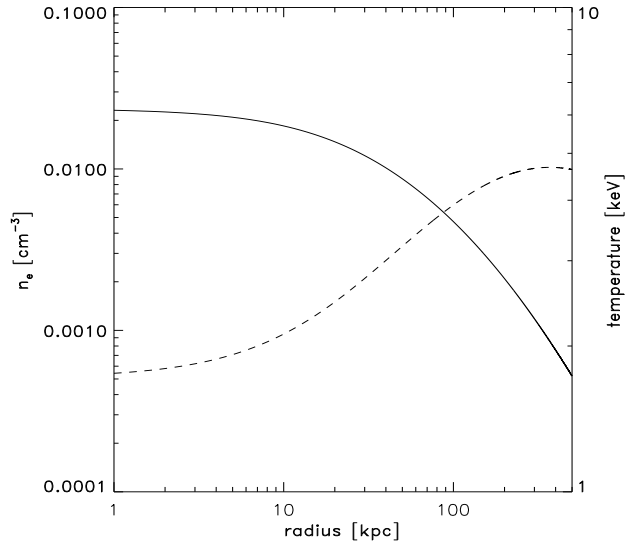}
\caption{Initial density (solid line) and temperature (dashed line) in the ICM.}
\label{plot1}
\end{figure}

\subsection{Magnetic fields}

We generated stochastic magnetic fields using a method similar to that described in
Ruszkowski et al. (2007). In brief, we set up magnetic fields in Fourier space and generate independent field components in
all three directions in ${\bf k}-$space. This is done to ensure that the the field has a random phase. For example, for the 
$x$-component, we set up a complex field such that 

\begin{eqnarray}
( {\rm Re}(B_{x}({\bf k})), {\rm Im}(B_{x}({\bf k}))) = (G(u_{1})B, G(u_{2})B),
\end{eqnarray}

\noindent
where 
$G$ is a function of a uniformly distributed independent random variables $u_{1}$ or $u_{2}$ that returns 
Gaussian-distributed random values. 
The amplitude $B$ is given by

\begin{eqnarray}
B\propto k^{-11/6}\exp\left[-\left(\frac{k}{k_{o}}\right)^{4}\right],
\end{eqnarray}

\noindent
where $k_{o}=2\pi/\lambda_{o}$ and $\lambda_{o}\sim 43h^{-1}$ kpc is a smoothing wavelength. 
We then apply a divergence cleaning operator in ${\bf k}-$space as follows:

\begin{eqnarray}
{\bf B}({\bf k}) \longrightarrow ({\bf 1}-\hat{\bf k}\hat{\bf k}){\bf B}({\bf k}),\\ \nonumber
\end{eqnarray}

\noindent
where $\hat{\bf k}$ is the unit vector in ${\bf k}$-space. 
Note that this operator does not change the shape of the power spectrum of the magnetic field fluctuations. 
This is sufficient for our purposes as we assume the field is dynamically unimportant.
We then use three-dimensional inverse fast Fourier transform to convert the field in ${\bf k}$-space to the real space.

\subsection{MHD equations with anisotropic thermal conduction and radiative cooling}

We solve the MHD equations, including field-aligned thermal conduction transport. 
These equations read:\\

\begin{eqnarray}
\frac{\partial\rho}{\partial t}+\nabla\cdot(\rho{\bf v}) = 0 
\end{eqnarray}
\begin{eqnarray}
\frac{\partial(\rho{\bf v})}{\partial t}+\nabla\cdot(\rho{\bf vv-BB})+\nabla p = \rho{\bf g} 
\end{eqnarray}
\begin{eqnarray}
\frac{\partial E}{\partial t}+\nabla\cdot({\bf v}(E+p)-{\bf B}({\bf v}\cdot{\bf B})) = \nonumber \\ 
\rho{\bf g}\cdot{\bf v} - \nabla\cdot{\bf F} + {\cal C}
\end{eqnarray}
\begin{eqnarray}
\frac{\partial{\bf B}}{\partial t} + \nabla\cdot ({\bf vB-Bv}) = 0, 
\end{eqnarray}

\noindent
where
\begin{eqnarray}
p=p_{\rm th}+\frac{B^{2}}{2}
\end{eqnarray}
\begin{eqnarray}
E=\frac{\rho v^{2}}{2}+\epsilon +\frac{B^{2}}{2},\\ \nonumber
\end{eqnarray}

\noindent
where $p_{\rm th}$ is the gas pressure and $\epsilon$ is the gas internal energy per unit volume. We assumed the
adiabatic index $\gamma = 5/3$. The anisotropic thermal conduction heat flux ${\bf F}$ is given by

\begin{eqnarray}
{\bf F}=-\kappa \hat{\bf e}_{B}(\hat{\bf e}_{B}\cdot{\mathbf\nabla} T),\\ \nonumber
\end{eqnarray}

\noindent
where $\hat{\bf e}_{B}$ is a unit vector pointing in the direction of the magnetic field and $\kappa$ is the Spitzer-Braginskii 
conduction coefficient given by $\kappa = 4.6\times 10^{-7}T^{5/2}$erg s$^{-1}$cm$^{-1}$K$^{-1}$.\\
\indent
In equation (21), ${\cal C}$ represents the cooling rate per unit volume. 
We use standard tabulated publicly available 
cooling curves \citep{sutherland93} for metallicity $Z = 0.3Z_{\odot}$.

\indent
Magnetic field evolution was solved by means of a directionally
unsplit staggered mesh algorithm (USM; Lee \& Deane 2009). 
The USM module is based on a finite-volume, high-order Godunov scheme combined with constrained
transport method (CT). This approach guarantees divergence-free magnetic field distribution. We implemented the anisotropic conduction unit
following the approach of Sharma \& Hammet (2007). More specifically, we applied monotonized central (MC) limiter to the conductive fluxes.
This method ensures that anisotropic conduction does not lead to negative temperatures in the presence of steep temperature 
gradients. Tests of the method are presented in accompanying paper (Ruszkowski et al. 2009, in preparation).

\subsection{Driven turbulence}

In order to emulate the effects of turbulence in the ICM, we employed a spectral forcing scheme that enables statistically stationary velocity fields
(Eswaran \& Pope 1988). This scheme utilizes an Ornstein-Uhlenbeck random process, analogous to Brownian motions in a viscous medium. At each timestep,
accelerations are applied to the gas. The acceleration field has vanishing mean and each acceleration mode has constant dispersion and is time-correlated.
That is, at each timestep the acceleration in a given direction is a Gaussian random variable with a fixed amplitude and decays with time as
$\propto \sqrt{1-f^{2}}$, where $f=\exp (-t/\tau_{\rm decay})$. This essentially means that the forcing term has a ``memory'' of the previous 
state. The phases are evolved in Fourier space and the divergence in the flow is cleaned making the 
flow incompressible, $\nabla \cdot v =0$. 
This is consistent with the Boussinesq approximation, and justified since the gas motions are significantly subsonic. 
Further details and the numerical tests of this method can be found in Fisher et al. (2008). The key quantity of interest is the velocity dispersion 
of the resultant velocity field $\sigma$ and its scaling with the injected energy and the injection spatial scale. 
This can be understood in terms of simple dimensional analysis
in an equilibrium situation when the turbulent driving is balanced by viscous dissipation in the fluid

\begin{eqnarray}
N_{\rm mode}\epsilon^{*}\sim\nu (\nabla\upsilon)^{2},\\ \nonumber
\end{eqnarray}

\noindent
where $N_{\rm mode}$ is the number of Fourier driving modes, $\epsilon^{*}$ is the energy injection per unit mass per mode and $\nu\sim L\upsilon$ 
is the gas viscosity for a a given largest eddy scale $L$. This implies the following scaling of the velocity dispersion 

\begin{eqnarray}
\sigma\sim\xi(N_{\rm mode}L)^{1/3}\epsilon^{*1/3},\;\; {\rm where} \;\; \xi\sim\mathcal{O}(1).\\ \nonumber
\end{eqnarray}

\noindent
We tested this scaling in a set of small box simulations and use it as a guide for choosing the parameters for the cluster simulations. 
\begin{table*}
 \centering
  \renewcommand{\thefootnote}{\thempfootnote} 
  \caption{Turbulence parameters of the performed runs}
  \label{steadytable}
  \begin{tabular}{@{}llllcccccc}
  \hline
  \hline
        & {$k_{\rm{min}}$} & {$k_{\rm{max}}$} & {$\tau_{\rm decay}$}   & $\epsilon^{*}$ & $\sigma$  \\
   Name & (cm$^{-1}$)      & (cm$^{-1}$)      & (s)                    & (cm$^{2}/$s$^{-3}$)             & (km/s)    \\
 \hline
 \hline
\\
weak    & $2.0\times 10^{-23}$  & $5.1\times 10^{-23}$ & 3.1$\times 10^{15}$ & $8.1\times 10^{-8}$ & $\sim$50  \\
strong  & $2.0\times 10^{-23}$  & $5.1\times 10^{-23}$ & 3.1$\times 10^{15}$ & $3.0\times 10^{-9}$ & $\sim$150  \\
\\
\hline
\label{table1}
\end{tabular}
\end{table*}
Table 1 provides a summary of the turbulence parameters. The values of the velocity dispersion quoted there are for the conductive case averaged
within $\sim100$ kpc from the cluster center. 
For closed box simulations, the velocity field is approximation Kolmogorov, though this is of course altered in a stratified medium.\\

\indent
The simulations were performed using the {\it FLASH} code (version 3.2). {\it FLASH} is a modular, parallel magnetohydrodynamic 
code that possesses adaptive mesh refinement capability. 
The three-dimensional computational domain was approximately 1 Mpc on each side, 
enclosing a large fraction of the cluster. 
The central regions of the cluster had an enhanced refinement level. The maximum spatial resolution for 6 levels of refinement was 
$\sim 2.7h^{-1}$ kpc. We performed a set of shorter simulations to verify that the results did not depend on the resolution for the 
chosen set of initial condition parameters. The simulations were performed on a 320-processor cluster located at the Michigan Academic Computing
Center at the University of Michigan in Ann Arbor and on the {\it Columbia} supercomputer at NASA Ames.\\

\begin{figure*}
\includegraphics[width=0.5\textwidth]{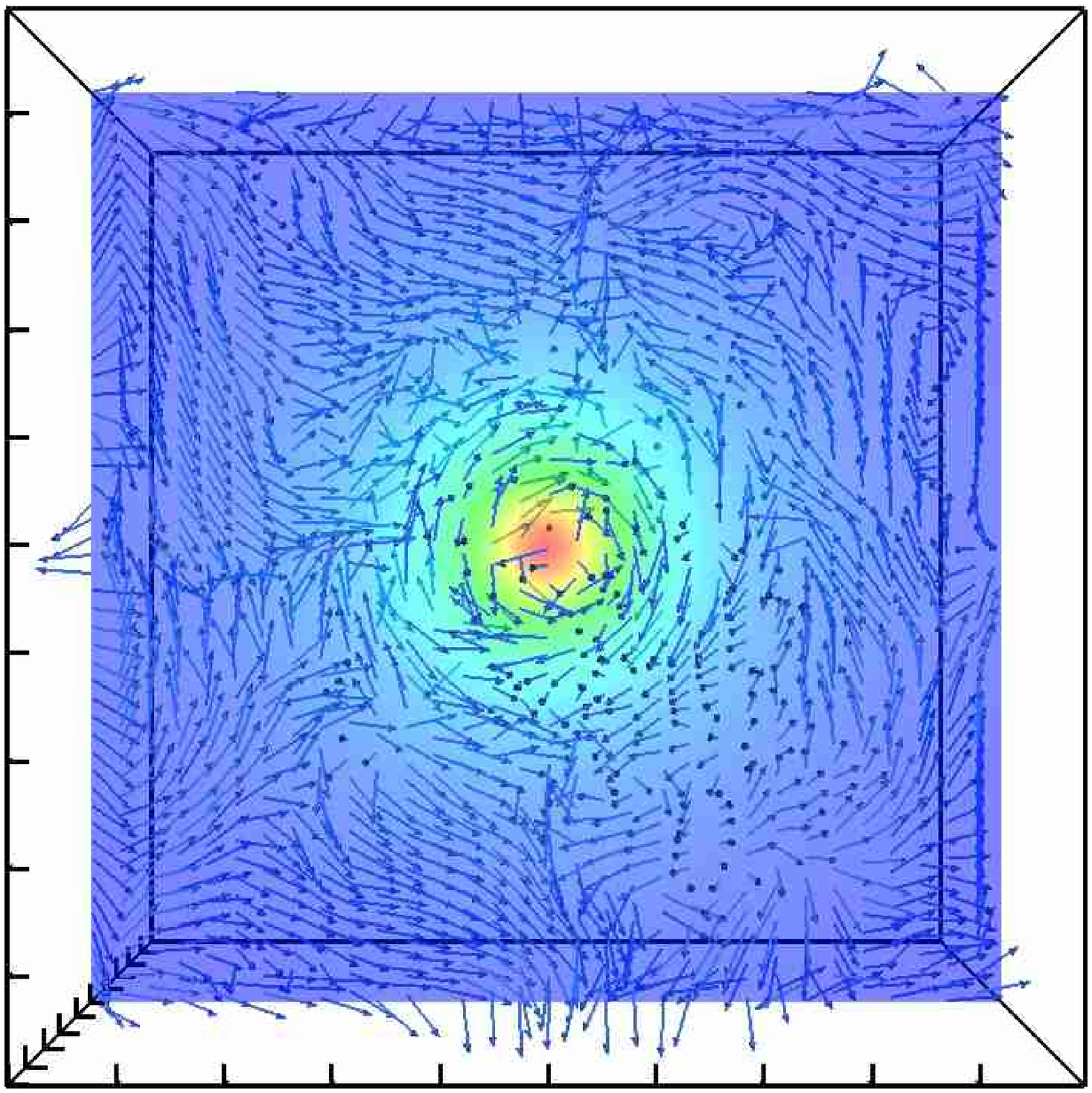}
\includegraphics[width=0.5\textwidth]{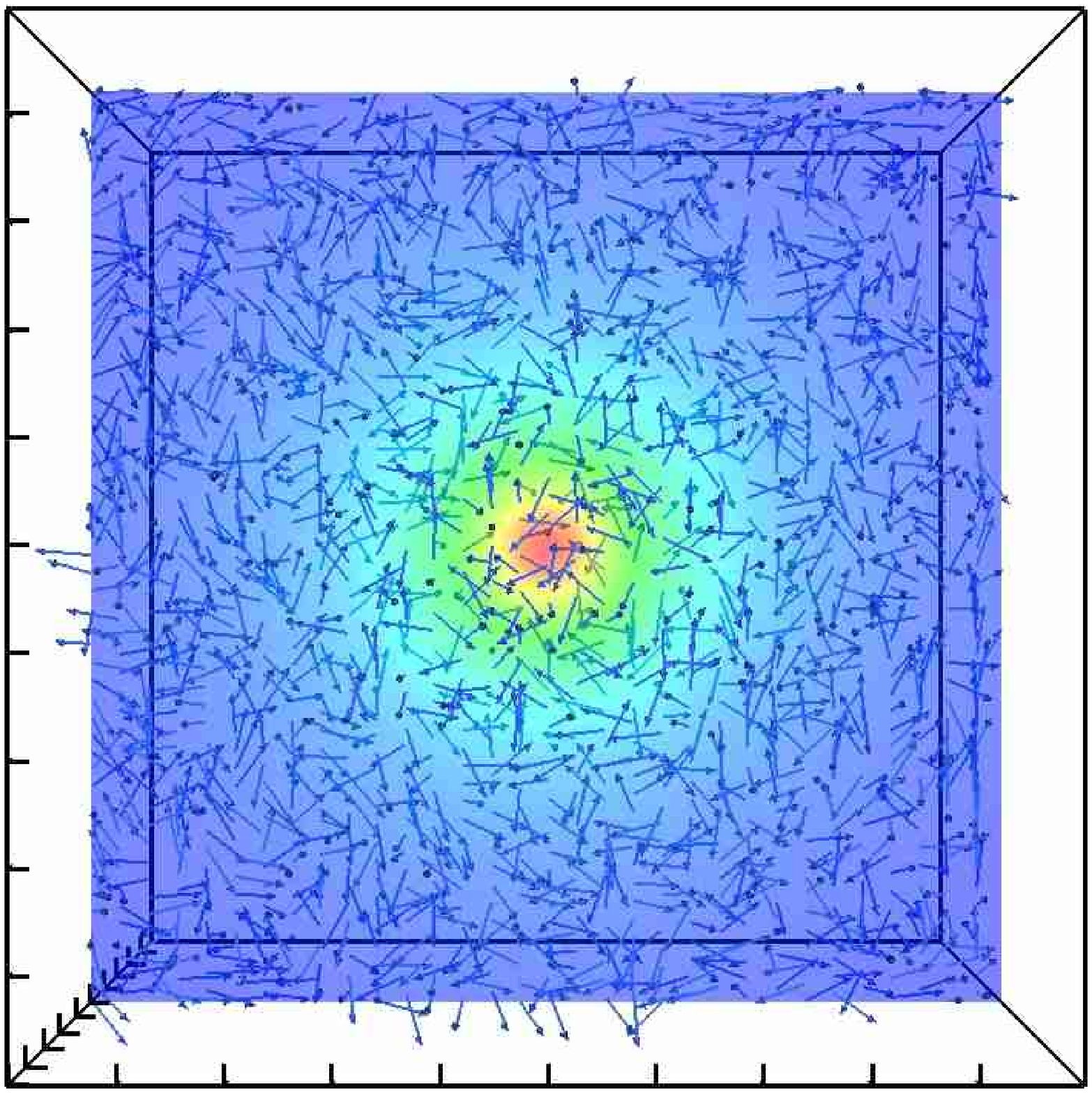}
\caption{Topology of the magnetic field (vectors) with superimposed gas density distribution (color background). The results are shown in the plane
intersecting the cluster center. The images are $\sim$1 Mpc on a side. Left panel corresponds to the end state of the simulation with anisotropic
conduction and the right panel to the simulation involving both the anisotropic conduction and very weak subsonic turbulence.}
 \label{plot1}
\end{figure*}

\section{Results}
\label{section:results}

In Figure 2 we present cross-sections through the cluster center. Both panels show the orientation of magnetic field 
vectors in the final state. Superposed on these vectors is the gas density distribution. Left hand side panel corresponds to 
the HBI case and the right one to the anisotropic conduction in the presence of weak stirring gas motions. In the latter case, 
the mean velocity dispersion is significantly subsonic and of the order of $\sim 50$ km/s.
It is evident from this figure that the addition of weak stirring ``outcompetes'' HBI and dramatically changes the topology of the field.
As explained below in detail, this randomization 
affects the effective level of thermal conduction.\\
\indent
In Figure 3 we quantify the effect of HBI and turbulence on the properties of the intracluster medium.
The parameters of all runs are summarized in Table 1 and their meaning is explained below.
All curves correspond to the quantities averaged within $\sim$100 kpc from the cluster center.
In the left panel we show the evolution of the anisotropy in the velocity field. The anisotropy is defined as
$\beta_{v}=1-\sigma^{2}_{t}/(2\sigma^{2}_{r})$, where $\sigma_{t}$ and $\sigma_{r}$ are the velocity dispersion 
in the tangential direction and radial directions, respectively. Thus, the isotropic case corresponds to 
$\beta_{v}=0$, while positive (negative) values of $\beta_{v}$ are for preferentially radial (tangential) gas motions. \\
\indent
We begin by discussing the pure HBI case (dashed lines) in all three panels in Figure 3.
In this case, the initial tangled magnetic field allows thermal conduction at 
the level of $\sim 1/3$ of the Spitzer value, and the resulting transfer of heat leads to radial flows to adjust to a new equilibrium. 
This rapidly shuts off once thermal conduction is suppressed. 
After the initial readjustment, the velocities become progressively
tangential (left panel). This is consistent with theoretical expectations for the low Froude number case, as in \S\ref{section:turbulence}; strong buoyancy restoring forces confine motion to be largely tangential. \\
\indent
In the middle panel we show
the anisotropy in magnetic field $\beta_{B}=1-\sigma^{2}_{Bt}/(2\sigma^{2}_{Br})$, where $\sigma_{Bt}$ and 
$\sigma_{Br}$ are the dispersion in the tangential and radial magnetic field components, respectively. This figure clearly illustrates 
that the HBI leads to strongly tangentially-biased magnetic fields. Finally, in the right panel we present the effective 
intrinsic thermal conduction coefficient of the ICM 
in terms of the Spitzer-Braginskii value. More specifically, 
the intrinsic effective conduction coefficient 
of the ICM is $f\kappa$, where $f$ is given by
$f=\langle F_{r}\rangle/\langle F_{r, \rm spitzer}\rangle$, where\\

\begin{eqnarray} 
\langle F_{r}\rangle \propto\langle T^{5/2}|(\hat{{\bf e}}_{T}\cdot\hat{{\bf e}}_{B})(\hat{{\bf e}}_{B}\cdot\hat{{\bf e}}_{r})|\rangle ,\\ \nonumber
\end{eqnarray}

\noindent
and \\

\begin{eqnarray} 
\langle F_{r, \rm spitzer}\rangle \propto\langle T^{5/2}|\hat{{\bf e}}_{T}\cdot\hat{{\bf e}}_{r}|\rangle .\\ \nonumber
\end{eqnarray}

\noindent
The volume averaging in the above expressions is performed within the central $\sim$100 kpc. The proportionality constant 
in Eq. 28 and Eq. 29 are identical. The versors ${\bf e}_{T}$, ${\bf e}_{B}$ and ${\bf e}_{r}$ point in the direction of the temperature gradient,
magnetic field vector and the radial direction, respectively. 
Note that the projection of the temperature gradient onto the B-field is given by the averaged absolute value, rather than a straight average. This correctly reflects the expectation that for 
$\hat{{\bf e}}_{T} = \hat{{\bf e}}_{r}$ and a completely randomized B-field, $f= \langle (\hat{{\bf e}}_{B}\cdot\hat{{\bf e}}_{r})^{2} \rangle =1/3$. We refer to $f$ as the Spitzer fraction below.
The effective conduction declines from the value close to the theoretical maximum $\sim 33\%$ to about $4\%$ of the Spitzer-Braginskii value.
The initial values of this ratio are somewhat lower then the theoretical value. This simply reflects the fact that we are sampling 
a finite volume of the cluster and so the actual value of the ratio depends on the random seed used to generate the field distribution.
The decline in the conductive flux is a consequence of the rearrangement of magnetic fields 
to become preferentially tangential. \\
\indent
We now compare the HBI reference results to simulations that include driven turbulence and anisotropic conduction.  
The solid black curve shows results from the weak turbulence driving case. The total energy injection due to stirring motions is
much smaller then the thermal energy content of the ICM. The gas velocity dispersion in the final state is $\sim 50$ km/s, i.e., much smaller the
the sound speed in the ICM. Such weak stirring motions are entirely conceivable and are consistent with, for example, repeated minor mergers, major mergers 
that stir the gas on large scales and cascade to smaller scales, AGN-driven motions or stirring by galaxy motions. 
For instance, \citet{nagai03} find from cosmological simulations that even relatively relaxed clusters have internal velocities $\sim 20\%$ of the internal sound speed; \citet{kim07} finds turbulent velocities $\sim 100-200 \, {\rm km \, s^{-1}}$ in numerical simulations of gravitational wakes of galaxies in clusters. 
The left panel demonstrates that the gas velocity field becomes preferentially tangential in this case, while 
the middle panel shows that the magnetic field is close to isotropic. This magnetic field topology is consistent with 
the right panel in Figure 2. The isotropization of the magnetic field leads to a significant boost in the level of the effective thermal conduction
as shown in the right panel of Figure 3. \\
\indent
In order to better understand the observed trends in the velocity, magnetic field and the effective 
conduction, we performed MHD runs with stirring but without conduction. An example of the result from one of these runs is shown as a solid purple line
in all three panels in Figure 3. It is evident that the stirring leads to preferentially tangential velocity field. This behavior is very similar to 
the case when both the stirring and HBI operate. The reason for the preferentially tangential gas motions in the absence of conduction 
is that the characteristic driving 
frequency is lower then the local Brunt-V\"{a}is\"{a}l\"{a} frequency $\omega_{BV}$. In this case, gas motions excite gravity waves ($g$-modes) 
that become trapped
within the radius where ``turbulence frequency''$\sim\sigma/\lambda\la\omega_{BV}$, where $\sigma$ and $\lambda$ are the gas velocity dispersion
and the characteristic eddy size, respectively. Consequently, the distribution of the orientation of magnetic field versors is also tangential (see middle 
panel). One subtle point worth emphasizing is that the magnetic fields have ``memory'' of the the fluid displacement history. In other words,
for weak fields, magnetic fields behave as ``tracer particles'' and, thus, the middle panel essentially quantifies the 
anisotropy in integrated fluid displacements. It is interesting to note that the magnetic field for weak stirring is close to isotropic in the 
conductive case, while it remains preferentially tangential in the absence of conduction. We argue that this effect can be qualitatively 
understood in terms of the magnitude of the buoyant restoring force. The magnitude of the restoring force in the magnetized medium with anisotropic
conduction depends on $\nabla T$ whereas for the pure MHD the restoring force depends on $\nabla S$, where $S$ is the gas entropy. The entropy gradient is steeper than the temperature gradient, so the buoyant restoring force in the conductive case is weaker
then in the pure MHD case. This means that it is easier to perturb and isotropize such a fluid. This may explain why the distribution of magnetic field
in a turbulent magnetized and conducting ICM is more isotropic than in the absence of conduction. We also note that 
the tangential bias has been seen in non-MHD adiabatic cosmological simulations that did not involve
thermal conduction (Rasia et al. 2004). In these simulations the characteristic velocity dispersion in the ICM was also highly subsonic and 
the volume-averaged velocity field anisotropy (coincidentally defined the same way as here) was $-2\la\beta\la 0$. This result is consistent
with  the idea of trapping of internal gravity waves.\\
\indent
The results for the strong stirring case are presented as solid blue (turbulence without thermal conduction) and solid green (turbulence with thermal conduction).
In these cases, the gas velocity dispersion is of the order of $\sigma\sim 150$ km/s (see next paragraph for the discussion of the velocity dispersion). 
Such motions, while being clearly subsonic, are nevertheless 
sufficiently vigorous to isotropize the velocity field with and without anisotropic conduction. They are also sufficiently powerful to overwhelm
the buoyant restoring force and make the magnetic field distribution isotropic in both cases. 
The Spitzer fraction is effectively 1/3 in this case (note that in the purely hydrodynamic case, the effective conduction is computed at the post-processing stage).\\
\indent
In Figure 4 we show the evolution of the volume-averaged velocity dispersions (left panel) and the profiles of velocity dispersions 
in the final state (right panel). As expected, the unperturbed HBI case leads to very small velocity dispersions whereas the volume-averaged 
velocity dispersion saturates
at $\sim 50$ km/s and $\sim 150$ km/s in the weak and strong stirring cases, respectively. 
The velocity dispersion profiles also show that the typical velocities are always significantly subsonic throughout the ICM.
Black and green lines are for the conductive case and purple and blue for the non-conductive ones.\\
\indent
In Figure 5 we present characteristic frequencies in the cluster: 
orbital (black), Brunt-V\"{a}is\"{a}l\"{a} (blue) $\omega_{\rm BV}$ for hydrodynamic case, $\omega_{\rm BV}^{\rm MHD}$  
(green) for MHD case, stirring frequency in the non-conductive case (yellow), stirring frequency in the MHD case (red). 
The stirring frequency is defined as $\sim\sigma/\lambda$, where $\lambda = 20$ kpc is some reference length of the order of the
coherence scale of the turbulent velocity field. Left panel is for the weak turbulence and the right panel for the strong turbulence case.
This figure is helpful in explaining the trends seen in the topology of the field and the Spitzer fraction (see Figure 3). 
Specifically, it can be seen that for the weak stirring and non-conductive case, the \brunt frequency exceeds the stirring frequency.
This implies trapping of internal gravity waves and tangential bias in the velocity and magnetic fields. This is consistent with the magnetic field 
topology in this case (purple curve, middle panel in Figure 3) 
that shows preferentially tangential magnetic fields. In the weak stirring conductive case, the 
stirring frequency is either comparable to the ``magnetic \brunt frequency'' $\omega_{\rm BV}^{\rm MHD}$ or it exceeds this frequency. This implies that 
it should be easier to randomize the magnetic fields. This is also consistent what is seen in Figure 3 (black curve, middle panel) that shows 
more isotropic magnetic fields in this case. The results for the corresponding strong stirring case are shown in the right panel of Figure 5.
In both the conductive and non-conductive cases, the stirring frequencies exceed $\omega_{\rm BV}$ and $\omega_{\rm BV}^{\rm MHD}$ which, consistent
with our findings, leads to isotropic magnetic field distribution.
\begin{figure*}
\includegraphics[width=0.33\textwidth]{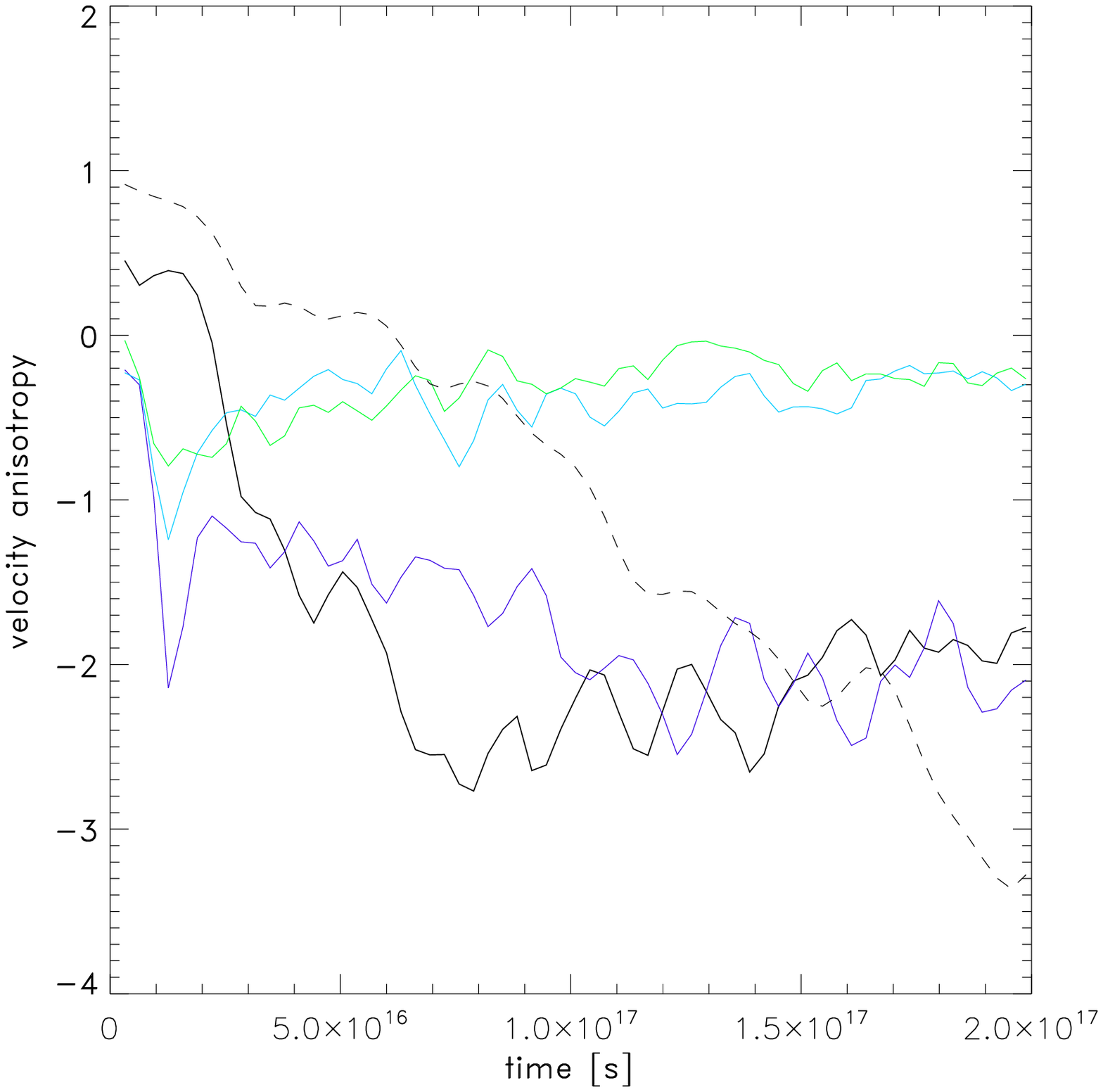}
\includegraphics[width=0.33\textwidth]{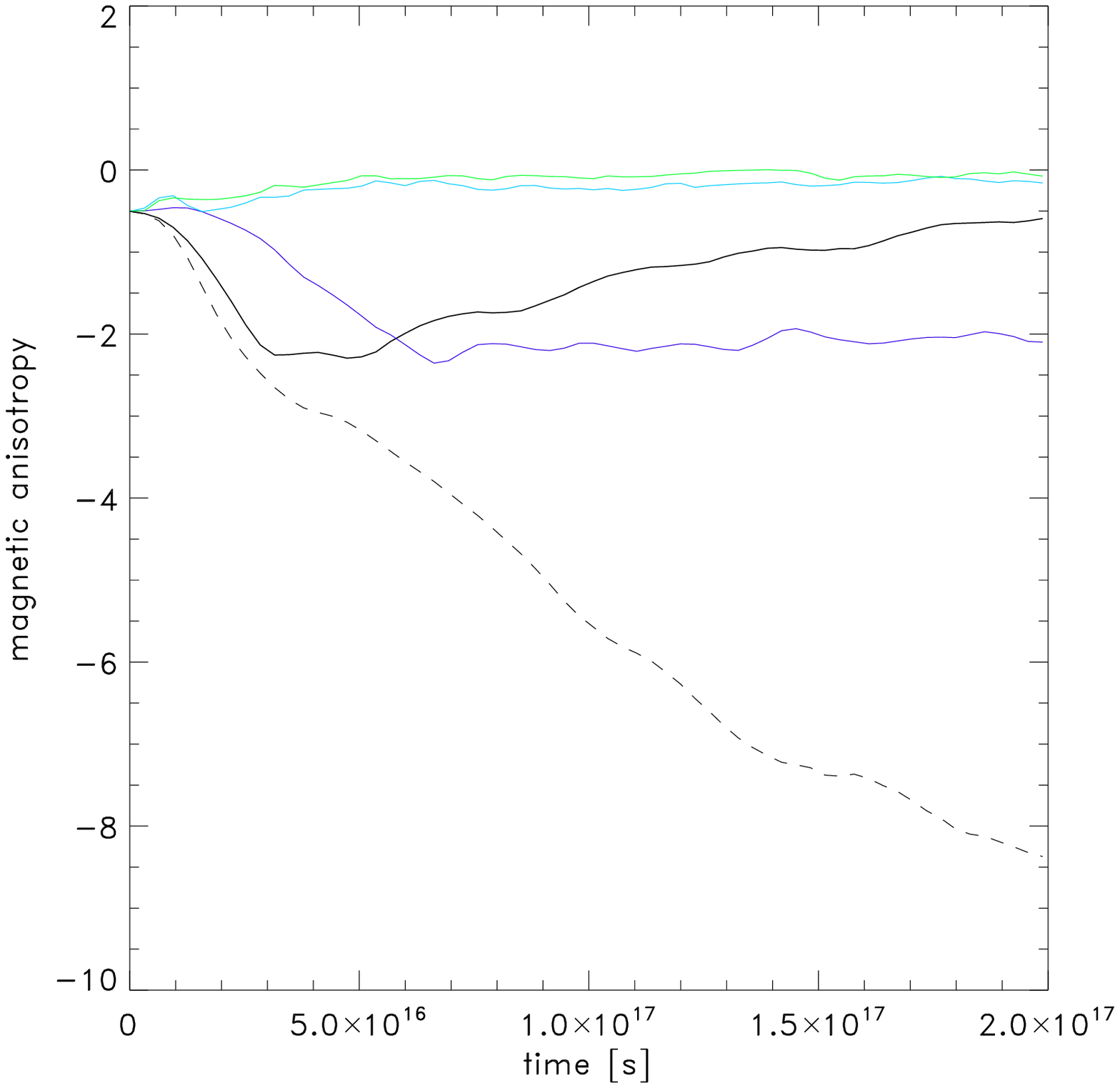}
\includegraphics[width=0.33\textwidth]{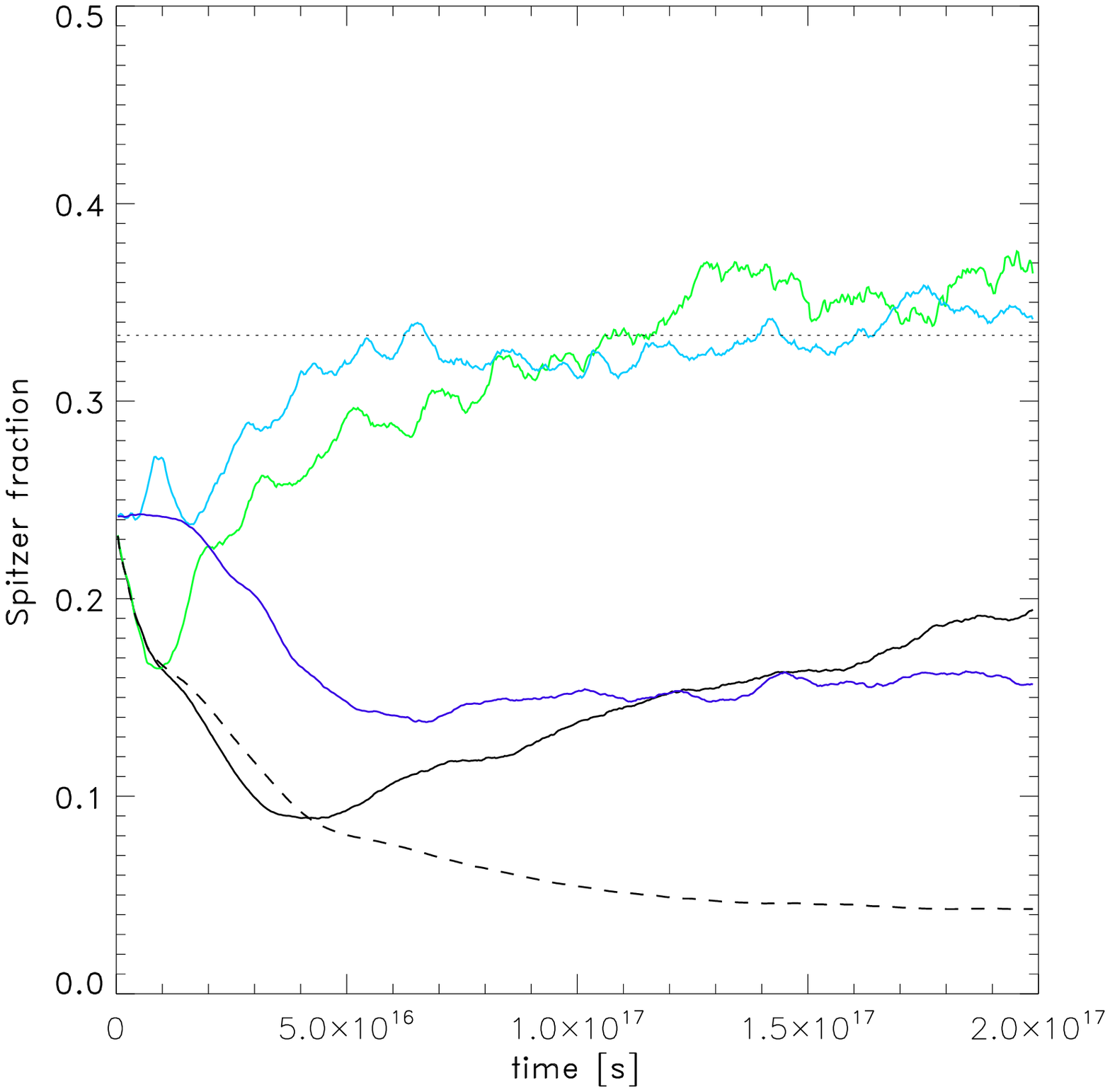} 
\caption{The evolution of 
velocity field anisotropy (left panel), magnetic field anisotropy (middle panel) and the effective thermal conduction as a fraction of the 
mean Spitzer-Braginskii conduction, all averaged within the central 100 kpc. Dashed lines correspond to the unbridled HBI instability. 
Solid black and purple lines are for weak turbulence with and without thermal conduction, respectively. Green and blue lines correspond to stronger turbulence with and without thermal conduction, respectively. See text for more details.}
 \label{plot2}
\end{figure*}
\begin{figure*}
\epsscale{1.0}
\includegraphics[width=0.5\textwidth]{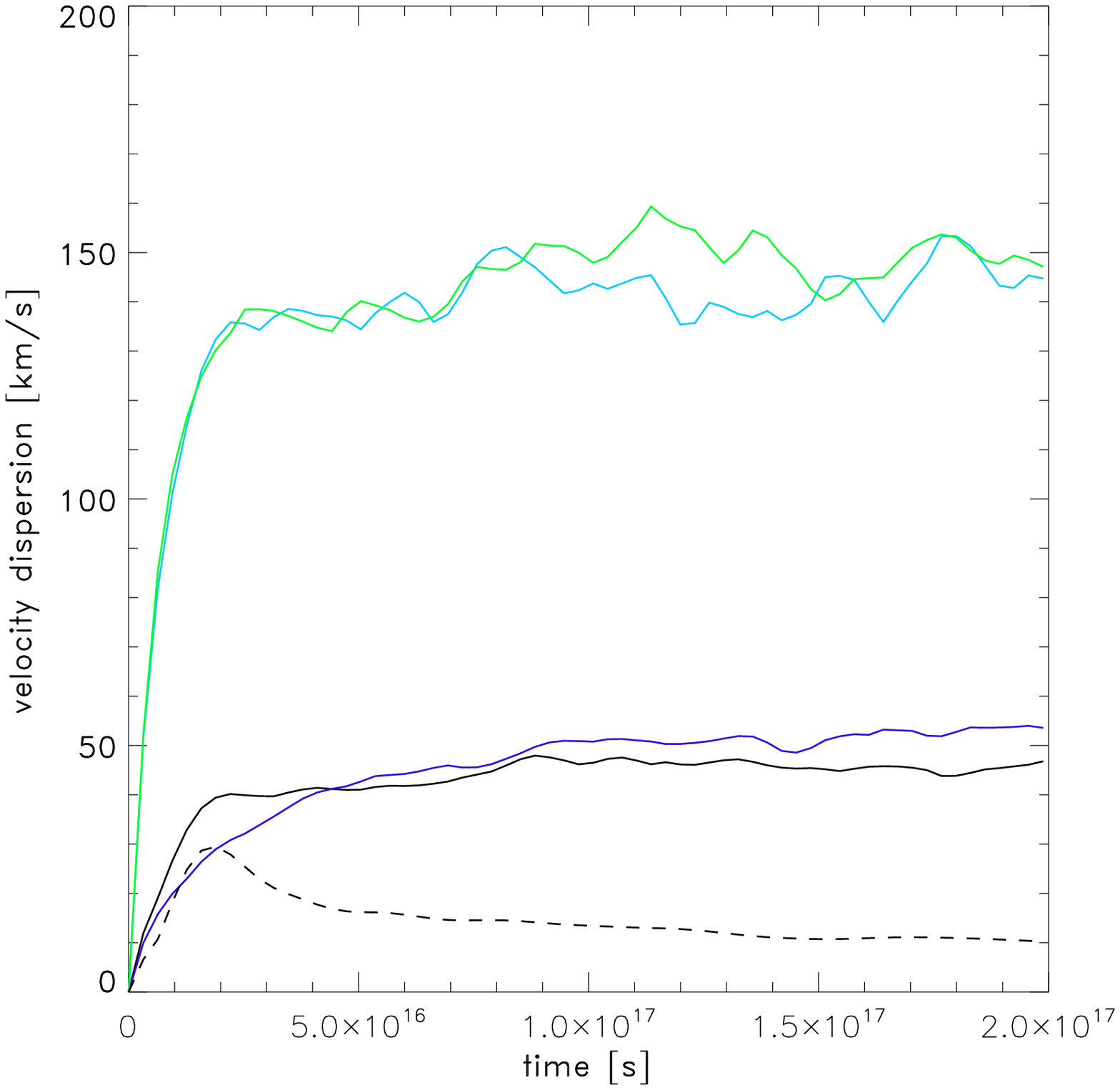}
\includegraphics[width=0.5\textwidth]{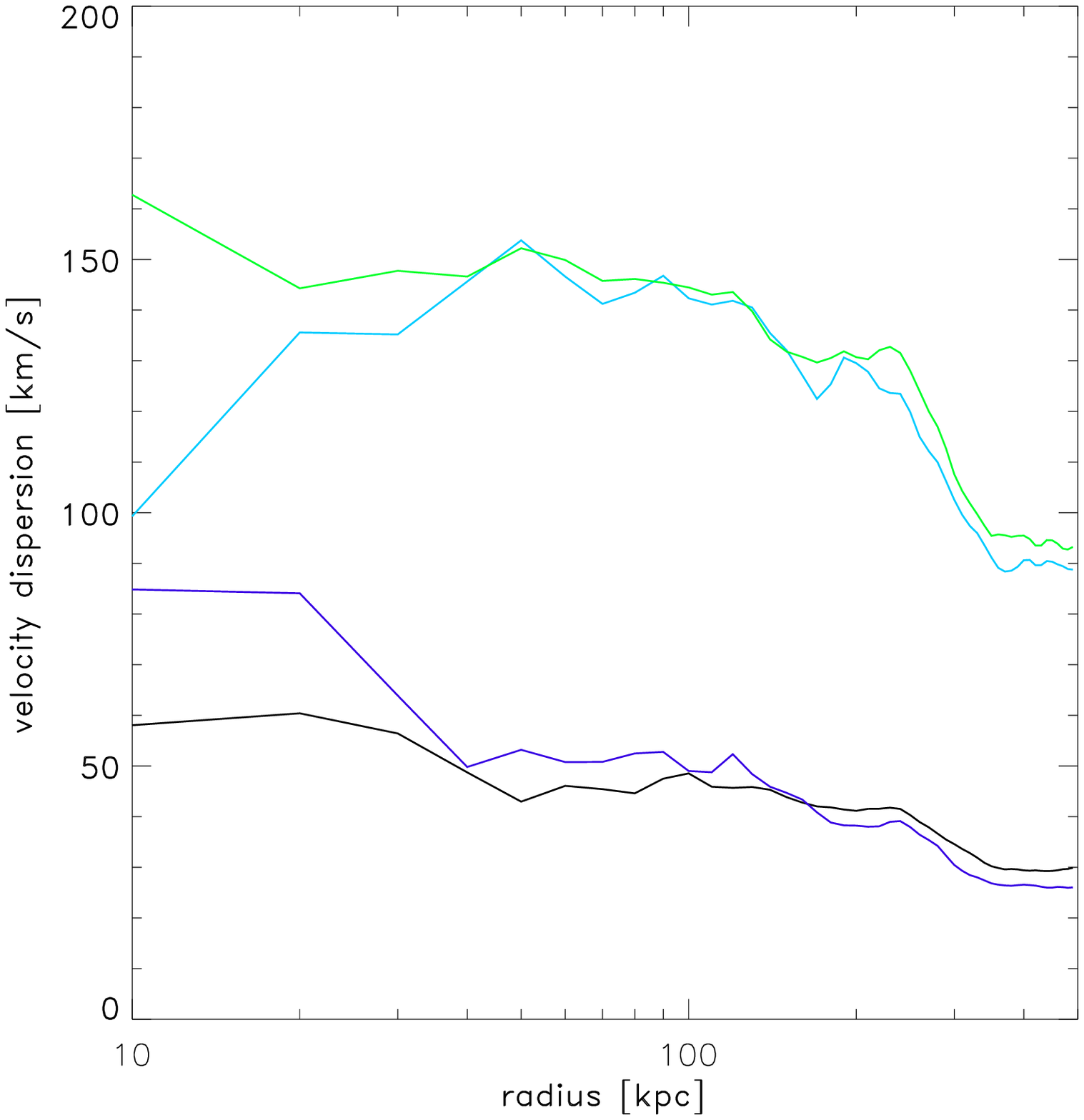}
\caption{Left panel: Evolution of the gas velocity dispersion in the central 100 kpc with time. Right panel: radial profile of velocity dispersion. 
The color coding of the lines is the same as in Figure 3.}
\label{plot3}
\end{figure*}
\begin{figure*}
\includegraphics[width=0.5\textwidth]{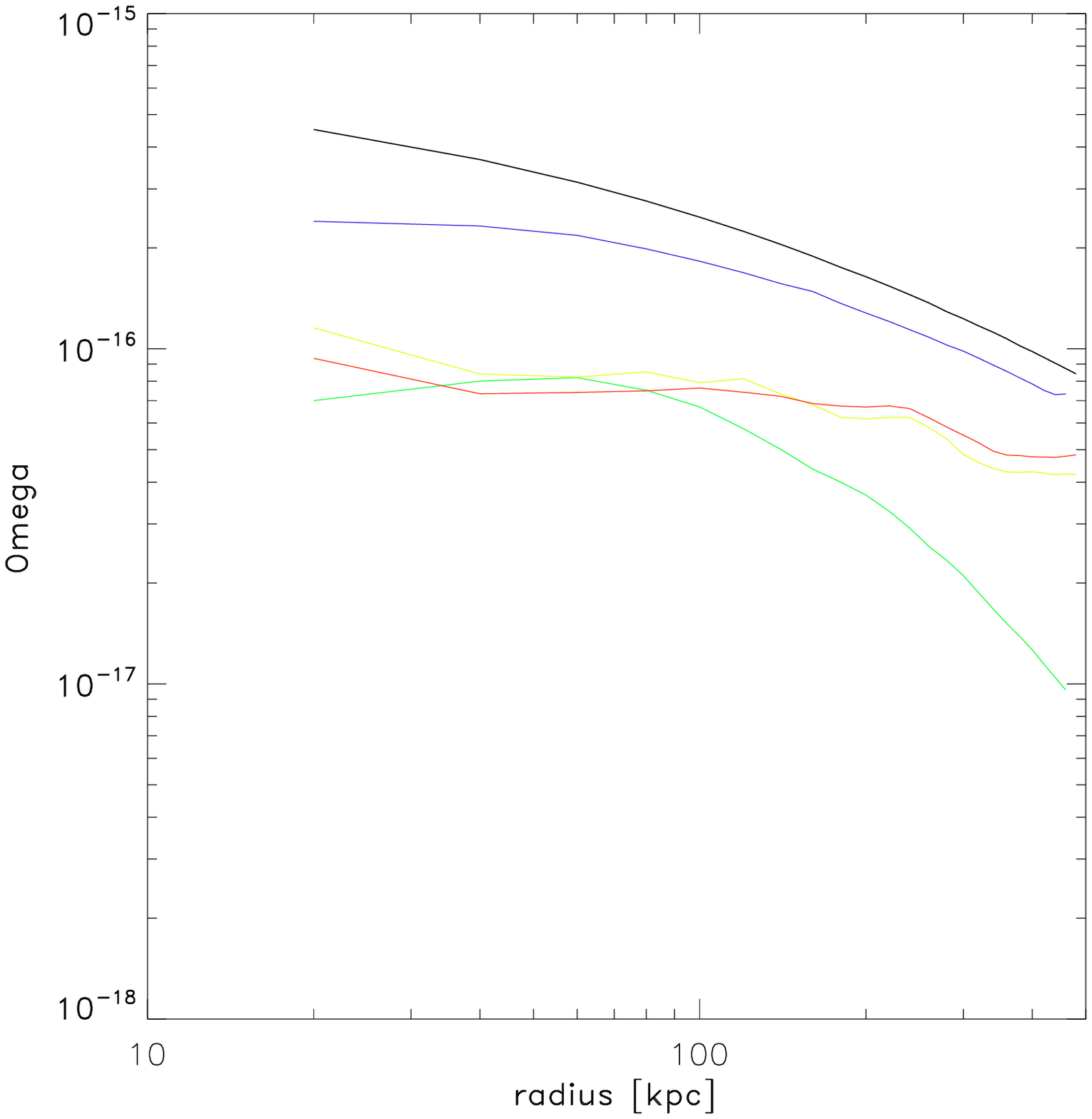}
\includegraphics[width=0.5\textwidth]{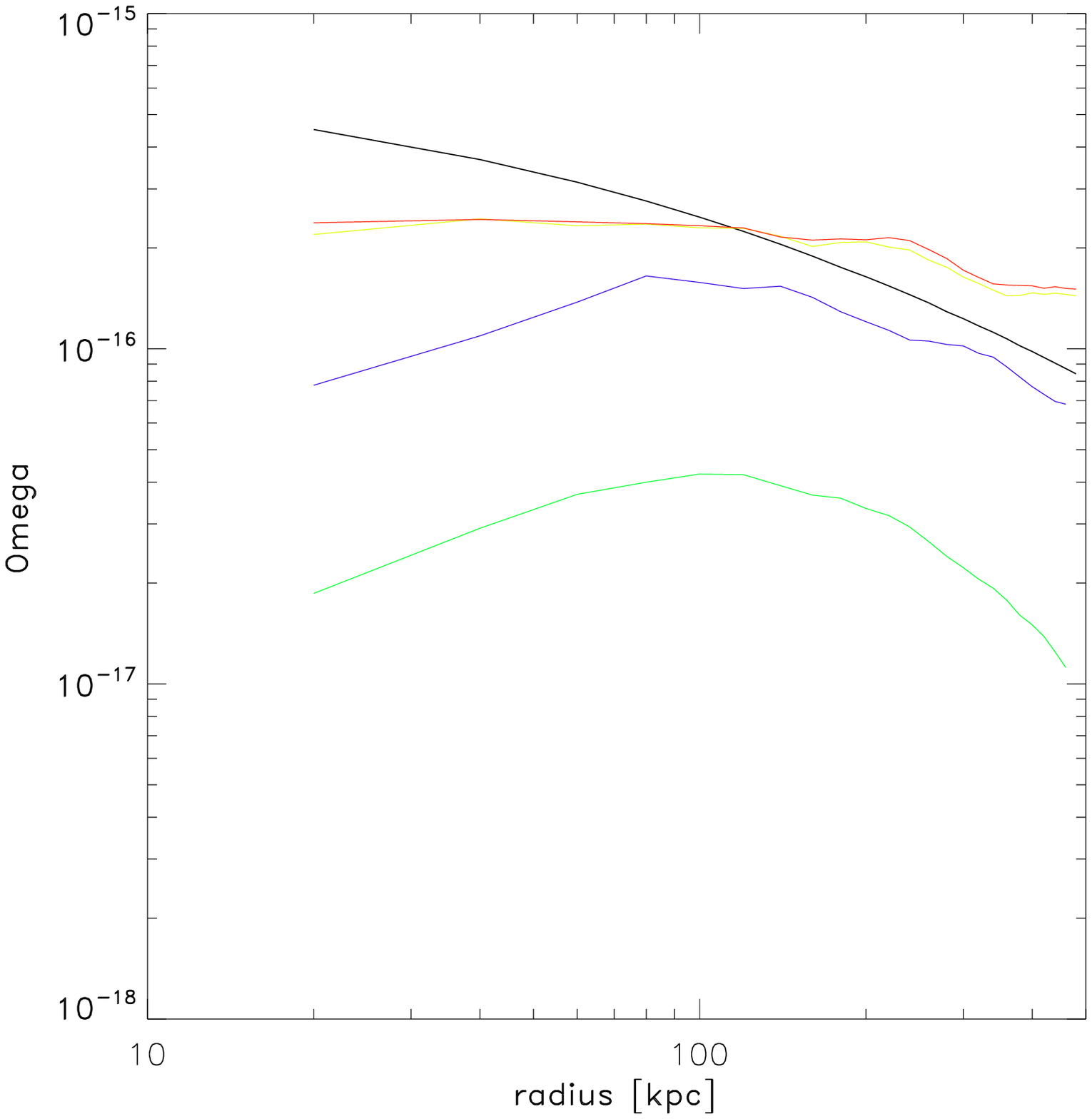}
\caption{Characteristic frequencies in the cluster (in [s$^{-1}$]): 
circular orbital (black), Brunt-V\"{a}is\"{a}l\"{a} (blue) $\omega_{\bf BV}$ for hydrodynamic case, $\omega_{\bf BV}^{\rm MHD}$  
(green) for MHD case, stirring frequency in the hydrodynamical case (yellow), stirring frequency in the MHD case (red). 
See text for definitions. Left panel is
for the weak turbulence and the right panel for the strong turbulence case.}
\label{plot4}
\end{figure*}
\begin{figure}
\includegraphics[width=0.5\textwidth]{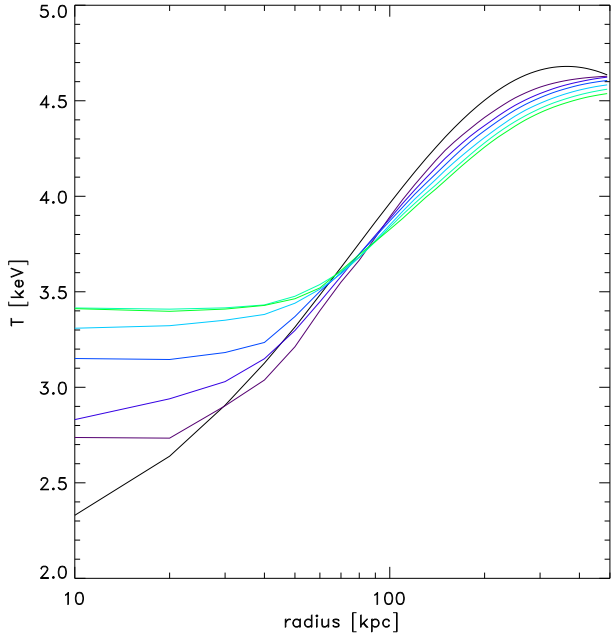}
\caption{Time sequence of temperature profiles for strong stirring with both thermal conduction and radiative cooling. The black curve denotes the initial state, whilst other curves are plotted every $\sim 1$ Gyr. Unlike the HBI case, no cooling catastrophe develops.}
 \label{plot4}
\end{figure}
\begin{figure}
\includegraphics[width=0.5\textwidth]{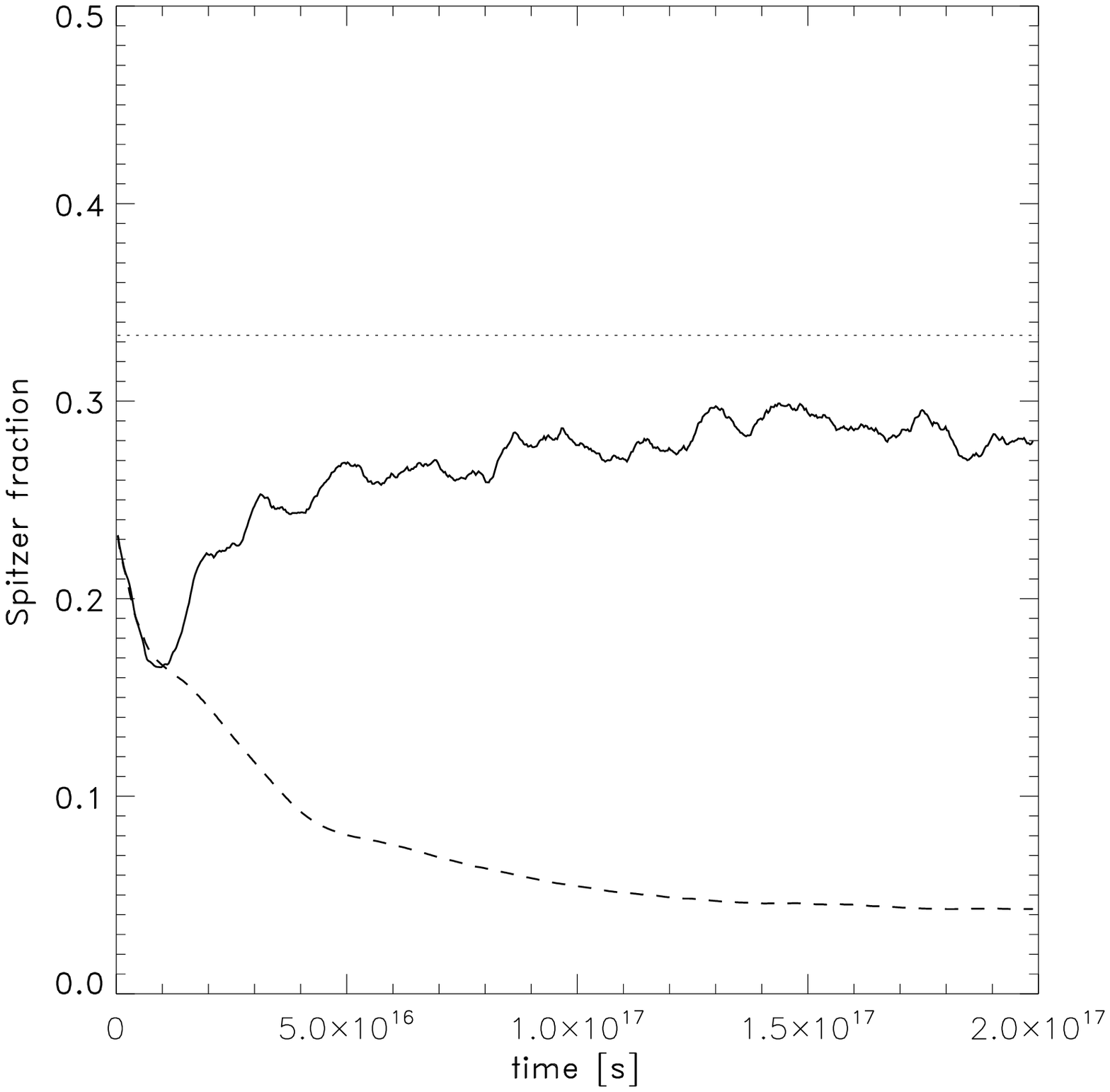}
\caption{Time evolution of the Spitzer fraction (solid line) for the same run as in Figure 6. 
Unlike in the HBI case (dashed line), 
the Spitzer fraction, even with cooling, approaches close to $f\sim 0.3$.}
 \label{plot4}
\end{figure}

\indent
We now discuss the runs with radiative cooling.
We performed a number of simulations that simultaneously include anisotropic thermal conduction, radiative cooling 
and different levels of stirring.
These runs were done for a range of gas densities and temperatures. We now present two cases for the initial conditions
described in Section 3 and then briefly comment on the cooling runs for different background ICM profiles.
As above, we consider exact same numerical setup as for the 
previous adiabatic cases but now add radiative cooling. As before, the weak stirring case has an extremely small turbulent 
gas velocity dispersion $\sigma \sim 50 \, {\rm km \, s^{-1}}$,
which is much smaller than the gas sound speed. In general, the presence of radiative cooling should not dramatically change the conclusions related 
to the topology of the magnetic field. The cooling catastrophe is delayed with respect to the initial cooling time by a factor of $\sim 4.3$.
Note that the energy in turbulent motions is much less than the thermal energy of the gas. This ratio is: 

\begin{eqnarray}
q\sim\left(\frac{\sigma_{\rm turb}}{\sigma_{\rm gas}} \right)^{2} \sim \frac{\gamma}{3} \mathcal{M}^{2}\\ \nonumber
\end{eqnarray}

\noindent
where $\mathcal{M}$ is the Mach number defined here as
$\mathcal{M}=\sigma/c_{s}$, and we have used $c_{s}^{2}=\gamma P/\rho = \gamma \sigma_{\rm gas}^{2}/3$. 
For example, for $T=3$ keV and $\sigma =50$ km/s (cf. Figure 4), we get $q\sim 1.4\times 10^{-3}$.
From Figure 4, we see that the turbulent energy is injected on a timescale $7\times 10^{8}$ years, 
which is comparable to the central cooling time.
Since $q\ll 1$, the turbulent energy injection rate is much smaller then the cooling rate.  
This estimate is independent of the presence of radiative cooling. The results for the strong stirring 
equivalent of the above run ($q \sim 1.3\times 10^{-2}$) show that the cooling catastrophe can be averted. 
The temperature profile flattens smoothly with time as a result of thermal conduction
and mixing (see Figure 6).\footnote{We point out that independent simulations involving radiative cooling
were performed by Parrish et al. (2009) who obtained qualitatively similar conclusions. In particular, they find that 
sufficiently strong stirring
can prevent excessive cooling. Our strong stirring case with cooling also confirms that finding.}\\
\indent
In simulations involving radiative cooling and turbulence resulting from 
acoustic-gravity waves, Fujita et al. (2004) found that cooling catastrophe can be prevented even in the absence of 
thermal conduction. However, the level of turbulence in those simulations was higher then in the cases that we consider.\\
\indent
Figure 7 is the analog of Figure 3 (right panel) and shows the behavior of the Spitzer fraction (solid line).
As in the adiabatic case, the ratio evolves asymptotically toward an approximately constant level but in this case the saturated level is slightly lower 
($\sim 30$\% of the Spitzer-Braginskii value). 
The unperturbed HBI case is shown for reference (dashed line).
We attribute the small offset between the theoretical maximum Spitzer fraction of 1/3 and the simulated one to the fact that the HBI is marginally 
more competitive in the cooling case, due to the steeper temperature profile. \\
\indent
We note that the cooling results described above come with possible caveats.
The above examples of cooling runs may not carry over to higher ICM densities. We performed a limited number of numerical experiments for 
higher ICM densities and found that the development of thermal instability is more likely in those cases. 
If so, it is possible that at higher ICM densities AGN may be required for stability.
This is consistent with the findings of
Conroy \& Ostriker (2008) who, based on 1D models, suggest that cooling catastrophe should occur at densities around 0.02 cm$^{-3}$
(somewhat below our initial central density, see Figure 1),
if no AGN feedback is present. 
The precise value of the critical density threshold (if confirmed in three-dimensional simulations) would have to be determined
by detailed numerical simulations. We defer this study to future work.

\section{Discussion}
\label{section:conclusions}

We have shown that a very low level of turbulent perturbations, $\sim 50-150 \, {\rm km/s}$ to the ICM --- entirely consistent with the expectations for cosmological infall, galaxy motions, mergers, or AGN activity --- can entirely alter the magnetic field distribution resulting from the HBI instability. Instead of preferentially tangential magnetic fields, the final field configuration can be easily randomized. As expected from simple theoretical considerations, this happens when the typical Froude number (equation \ref{eq:Fr_MHD}) $Fr^{\rm MHD} \gsim \mathcal{O}(1)$. 
Note that weak, low frequency perturbing motions lead to trapped $g$-modes and result in preferentially tangential gas motions, due to strong restoring buoyancy forces in the vertical direction; thus, magnetic fields could in principle become tangential even in the absence of thermal conduction and the HBI. However, this tangential velocity bias is also lifted for $Fr_{\rm MHD} \gsim \mathcal{O}(1)$. This has several immediate consequences:
\begin{itemize}
\item{The boost in thermal conduction restores the heat flow from the hot outer layers to the central cool core. This is crucial to averting massive overcooling and heating the cluster cores in a stable fashion.
Indeed, our runs with radiative cooling showed that a cooling catastrophe was averted even in the absence of AGN heating. 
Thus, naive models which assume thermal conductivity with Spitzer fraction $f\sim 1/3$ may perhaps be reasonably applied to clusters after all.
Since conduction is regulated by turbulence, it will in general be time-dependent. A sudden boost in thermal conduction can modulate a transition between a cool core (CC) to a non cool core (NCC) cluster (\citet{guo09}; we also see some hints of this in Fig \ref{plot4}). Thus, mergers, while generally insufficient in the pure hydrodynamic case to effect a CC to NCC metamorphosis \citep{poole08}, may be effective once the boost in thermal conduction due to additional turbulence is taken into account. If AGN are the main source of turbulence, this implies feedback between accretion onto the AGN and thermal conduction. Thus, AGN could exert another level of self-regulation and thermostatic control beyond straight kinetic heating. 
}
\item{Our predictions might be testable in the future: magnetic topology can be measured by radio polarization measurements (e.g., Pfrommer \& Dursi 2009), while turbulence can be measured by a high spectral-resolution calorimeter. Alternatively, our predictions can be used to provide indirect constraints on turbulence. For instance, the absence of large-scale ordered fields would require a minimal level of turbulence, and argue against a fully relaxed ICM. The resulting turbulent pressure support affects mass estimates which assume hydrostatic equilibrium \citep{lau09}, and the use of clusters for cosmology.}
\item{The distribution of metals in clusters is known to be broader than that of the stars. As shown by \citet{sharma09,sharma09a}, metal mixing in a stratified plasma is much more effective once conduction is at play, because of the weaker nature of restoring buoyancy forces. This allows broad metallicity profiles to be obtained without inversion of the central entropy profile (which is not observed). The restoration of thermal conduction by stirring motions greatly aids this process. In the limit where conduction is so efficient that the cluster becomes isothermal (so that $\omega_{\rm BV}^{\rm HBI} \rightarrow 0$)
or conduction becomes isotropic, fluid elements have the same temperature (and hence density) as their surrounding, 
and fluid elements become neutrally buoyant. This makes metal mixing maximally efficient. We note that any mechanism which disperses metals over wide distances (e.g., turbulent diffusion, Rebusco et al. 2008) generally excites turbulent velocities sufficient to isotropize the magnetic field. Thus, the relatively broad metallicity profiles in clusters is consistent with non-negligible stirring.}
\end{itemize}

Our stirring methods ensure by construction that the ensuing turbulence is volume-filling. However, this is by no means guaranteed in nature. For instance, galactic wakes will be relatively narrow, with a cross-section of order the galaxy size; similarly, rising AGN bubbles will not induce velocity fluctuations over a large solid angle. A straightforward way to ensure volume-filling turbulence, as is required to stem the HBI, is to excite trapped $g$-modes which are repeatedly reflected and focused within a resonance region where $\omega < \omega_{\rm BV}^{\rm MHD}$, as discussed in \S\ref{section:turbulence}. At the same time, $\omega > \omega_{\rm BV}^{\rm MHD}$ is required to overwhelm buoyancy forces, an apparently contradictory requirement. Of course, the assumption of a single frequency is too simplistic: stirring motions contain many harmonics and will be scale dependent, generally increasing in frequency as turbulence cascades toward smaller scales. Thus, large scale $g$-modes could fall below the \brunt frequency, ensuring trapping and amplification, but the resulting small scale turbulence could potentially still have turnover frequencies exceeding $\omega_{\rm BV}^{\rm MHD}$. We will examine this in future work. 

Although we have focused on the HBI in this paper, the same Froude number considerations apply with equal force to the MTI unstable outer regions of the cluster, where temperature falls with radius and the MTI causes field lines to become preferentially radial. Here, if $Fr^{\rm MHD} > \mathcal{O}(1)$, turbulent motions can also overwhelm buoyant forces and randomize the field. Recently, there has been tantalizing evidence from the polarized emission surrounding the magnetic drapes of galaxies sweeping up magnetic fields in the Virgo cluster, that outside a central region, the magnetic field might be preferentially oriented radially \citep{pfrommer09}, as one might expect from the MTI. If this result continues to hold up, this would imply the Froude number is below the critical value at these radii (thus providing an indirect constraint on turbulent velocities), or additional physical processes, not modeled here, are at play. \\

\section*{Acknowledgments}
The software used in this work was in part developed by the DOE-supported ASC/Alliance Center for
Astrophysical Thermonuclear Flashes at the University of Chicago.
MR thanks Jeremy Hallum for his invaluable help with maintaining the computing cluster at the Michigan Academic Computing Center where most of the 
computations were performed. MR thanks Justin Nieusma for technical assistance with performing the simulations.
We thank Dongwook Lee, Ian Parrish, Eliot Quataert, Elena Rasia, Prateek Sharma and Min-Su Shin 
for discussions. We would like to point out that conclusions from cooling runs, similar to the ones presented here, were obtained independently by
Ian Parrish and collaborators. We are grateful to this group for sharing some of their results.
SPO acknowledges support by NASA grant NNG06GH95G, and NSF grant 0908480. MR acknowledges support by {\it Chandra} theory grant TM8-9011X.
MR and SPO thank Institute of Astronomy, Cambridge, UK and Max Planck Institute for Astrophysics, Garching, Germany for their hospitality.

\bibliography{master_references} 

\label{lastpage}

\end{document}